\documentclass[11pt]{article}
\usepackage{amsmath}
\usepackage{amsfonts}
\usepackage{amssymb}
\usepackage{graphicx}

\def\bea{\begin{eqnarray}}
\def\eea{\end{eqnarray}}

\begin{document}
\begin{center}
\LARGE {\bf The effect of decoherence on mixing time in
continuous-time quantum walks on one-dimension regular networks
 }
\end{center}
\begin{center}
{\bf S. Salimi {\footnote {E-mail: shsalimi@uok.ac.ir}}, R. Radgohar {\footnote {E-mail: r.radgohar@uok.ac.ir}}}\\
 {\it Faculty of Science,  Department of Physics, University of Kurdistan, Pasdaran Ave., Sanandaj, Iran} \\
 \end{center}
\vskip 3cm
\begin{center}
{\bf{Abstract}}
\end{center}
In this paper, we study decoherence in continuous-time quantum walks
(CTQWs) on one-dimension regular networks. For this purpose, we
assume that every node is represented by a quantum dot continuously
monitored by an individual point contact(Gurvitz's model). This
measuring process induces decoherence. We focus on small rates of
decoherence and then obtain the mixing time bound of the CTQWs on
one-dimension regular network which its distance parameter is $l\geq
2$. Our results show that the mixing time is inversely proportional
to rate of decoherence which is in agreement with the mentioned
results for cycles in~\cite{FST,VKR}. Also, the same result is
provided in~\cite{SSRR} for long-range interacting cycles. Moreover,
we find that this quantity is independent of distance parameter
$l(l\geq 2)$ and that the small values of decoherence make short the
mixing time on these networks.

\newpage

\section{Introduction}
Quantum walk(QW) as a generalization of random walk(RW) is
attracting great attention in many research areas, ranging from
solid-state physics~\cite{GHW} to quantum computing~\cite{MANC}.
Experimental implementations for the quantum walks have been
presented in~\cite{RLBL, BCS, PZ}. In recent years, two types of the
quantum walks exist in the literature: the continuous-time quantum
walks(CTQWs)~\cite{AMB, JB, JSS, NKLT, JSTA} and the discrete-time
quantum walks(DTQWs) ~\cite{CHKS, HK, CSL, NK, JSAM, FH}. The
relationship between the CTQWs and the DTQWs has been considered
in~\cite{MC, DA, KP}. The CTQWs have been studied on star
graph~\cite{SSS,xp1}, on direct product of cayley
graphs~\cite{SJDP}, on quotient graphs~\cite{SSQG}, on odd
graphs~\cite{SSOG}, on trees~\cite{NKT} and on ultrametric
spaces~\cite{NKUS}. All of these articles have focused on the
coherent CTQWs. The effect of decoherence in the CTQWs has been
studied on hypercube~\cite{FWS, AR}, on cycle~\cite{FST}, on
line~\cite{RSAAD, KTL}, on $N$-cycle~\cite{KTNC} and on long-range
interaction cycles~\cite{SSRR}. Here, we study the CTQWs on
one-dimension$(1D)$ networks with diatance parameter $l\geq 2$ which
can be constructed as follows~\cite{XPXR}: we construct an one
dimensional ring lattice of $N$ nodes, each node of which is
connected to its 2$l$ nearest neighbors($l$ on either side). The
structure of one-dimension regular network with $N=8$ and $l=3$ is
illustrated in Fig. 1. One-dimension regular networks have broad
applications in various coupled systems, for example, Josephson
junction arrays~\cite{KW}, small-world networks~\cite{MPB} and
synchronization~\cite{BBH}. In our paper, the network nodes are
represented by identical tunnel-coupled quantum dots(QDs). The walks
are performed by an electron initially placed in one of the dots. An
individual ballistic one-dimension point-contact is placed near each
dot as "detector" which its resistance is very sensitive to the
electrostatic field generated by electron occupying the measured
quantum dot. Decoherence is induced by continuous monitoring of each
network node with nearby point contact. We focus on small rates of
decoherence, then calculate the probability distribution and the
mixing time bound of the CTQWs on one-dimension regular network with
distance parameter $l\geq 2$. Our analytical results show that small
decoherence can make short the mixing time in the CTQWs. The same
result was produced for cycles in~\cite{FST} and for long-range
interacting cycles in~\cite{SSRR}. Moreover, we show that for small
rates of decoherence, the mixing time is independent of distance
parameter $l(l\geq 2)$. \clearpage
\begin{figure}[h]
\vspace{.5cm}\includegraphics{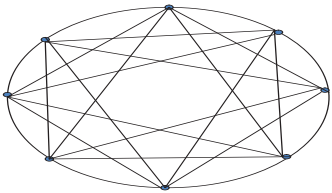}
 \caption{One-dimension regular network with $N=8$ and
$l=3$}
\end{figure}

This paper is organized as follows:
\\ In Sec. 2, we briefly review the properties of
CTQWs on one-dimension regular networks. In Sec. 3, we study the
decoherent CTQWs on the underlying network. We assume that the rate
of decoherence is small and obtain the probability distribution,
analytically in Sec. 4. The bound of the mixing time and its
physical interpretation are provided in Sec. 5. Conclusions and
discussions are given in the last part, Sec. 6.

\section{CTQWs on $1D$ regular network}

The properties of network is well characterized by the spectrum of
adjacency matrix of associated graph. The network adjacency
matrix($A$) is defined in the following way: $A_{ij}=1$ if nodes $i$
and $j$ are connected and otherwise $A_{ij}=0$. The Laplacian is
defined as $L=A-D$, where $D$ is a diagonal matrix and $D_{j,j}$ is
the degree of vertex $j$. Classically, the continuous-time random
walks(CTRWs) are described by the master equation~\cite{GHW, NVK}

\begin{eqnarray}\label{1}
  \frac{d}{dt}p_{k,j}(t)=\sum_{l}T_{kl}p_{l,j}(t),
   \end{eqnarray}
where $p_{k,j}(t)$ is the conditional probability to find the walker
at time $t$ and node $k$ when starting at node $j$. The transfer
matrix of the walk, $T$, is related to the adjacency matrix by
$T=-\gamma L$. For the sake of simplicity, we assume that the
transmission rate $\gamma$ of all bonds to be equal. The formal
solution of Eq. (1) is

\begin{eqnarray}\label{2}
   p_{k,j}(t)=\langle k|e^{Tt}|j\rangle.
   \end{eqnarray}
The quantum-mechanical extension of the CTRW is called the
continuous-time quantum walk(CTQW). The CTQW is obtained by
replacing the Hamiltonian of system with the classical transfer
operator, $H=-T$~\cite{GHW, NVK, EFSG}.
\\The Hamiltonian matrix $H$ for one-dimension regular network is written as the following form~\cite{XPXR}

\begin{eqnarray}\label{3}
   H_{ij}=\langle i|H|j \rangle=\left\{
                                \begin{array}{ll}
                                  -2\emph{m}, & \hbox{if $i=j$;} \\
                                  +1, & \hbox{if $i=j\pm z, z\in[1,\emph{l}]$;} \\
                                  0, & \hbox{Otherwise,}
                                \end{array}
                              \right.
        \end{eqnarray}
that the basis vectors $|j\rangle$ associated with the nodes $j$
span the whole accessible Hilbert space. In these basis, the
Schr\"{o}dinger equation(SE) is

\begin{eqnarray}\label{4}
   i\frac{d}{dt}|j\rangle=H|j\rangle,
   \end{eqnarray}
where we set $m\equiv 1$ and $\hbar\equiv 1$. The Hamiltonian acting
on the state $|j\rangle$ can be written as

\begin{eqnarray}\label{5}
   H|j\rangle=-(2l+1)|j\rangle+\sum_{z=-l}^{l}|j+z\rangle, z\in
   Integers
   \end{eqnarray}
which is the discrete version of the Hamiltonian for a free particle
moving on a lattice. It is well known in solid state physics that
the solutions of the SE for a particle moving freely in a regular
potential are Bloch functions~\cite{OMAB, JMZ}. Thus, the time
independent SE is given by

\begin{eqnarray}\label{6}
H|\Phi_{n}\rangle=E_{n}|\Phi_{n}\rangle,
   \end{eqnarray}
where eigenstates $|\Phi_{n}\rangle$ are Bloch states. The periodic
boundary conditions require that $\Phi_{n}(N)=\Phi_{n}(0)$, where
$\Phi_{n}(x)=\langle x|\Phi_{n}\rangle$. This restricts the
$\theta$-values to $\theta=\frac{2\pi n}{N}$, where $n=0,1,\ldots
,N-1$. The Bloch state $|\Phi_{n}\rangle$ can be expressed as a
linear combination of the states $|j\rangle$ localized at nodes $j$,

\begin{eqnarray}\label{7}
   |\Phi_{n}\rangle=\frac{1}{\sqrt{N}}\sum_{j=0}^{N-1}e^{-i\theta_{n}
   j}|j\rangle.
   \end{eqnarray}
Substituting Eqs. (5) and (7) into Eq. (6) , we obtain the
eigenvalues of the system as

\begin{eqnarray}\label{8}
   E_{n}=-2l+2\sum_{j=1}^{l}cos(j\theta_{n}).
   \end{eqnarray}
The time evolution of state $|j\rangle$ starting at time $t_{0}$ is
given by $U(t,t_{0})|j\rangle$, where
$U(t,t_{0})=\exp(-iH(t-t_{0}))$ is the quantum mechanical time
evolution operator. Hence, the transition amplitude
$\alpha_{k,j}(t)$ from state $|j\rangle$ at time $0$ to state
$|k\rangle$ at time $t$ is

\begin{eqnarray}\label{9}
   \alpha_{k,j}(t)=\langle k|e^{-iHt}|j\rangle.
   \end{eqnarray}
Applying $E_{n}$ and $|\Phi_{n}\rangle$ to represent the
$\emph{n}$th eigenvalue and eigenvector of $H$, the classical and
quantum transition probabilities between two nodes can be written as

\begin{eqnarray}\label{10}
   P_{k,j}(t)=\sum_{n}e^{-tE_{n}}\langle k|\Phi_{n}\rangle\langle
   \Phi_{n}|j\rangle,
   \end{eqnarray}

\begin{eqnarray}\label{11}
   \pi_{k,j}(t)=|\alpha_{k,j}(t)|^{2}=|\sum_{n}e^{-itE_{n}}\langle k|\Phi_{n}\rangle\langle \Phi_{n}|j\rangle|^{2}.
   \end{eqnarray}

\section{The Decoherent CTQWs on $1D$ regular network}
In this section, we investigate decoherence induced by the point
contact(PC) detector measuring the occupation of one of the quantum
dots(QDs) in a double-dot system. The measurement process is shown
schematically in Fig. 2. We assume all electrons to be spin-less
fermions and the tunneling between the PCs and the QDs to be
negligible, but we take into account Coulomb interaction between
electrons in the QD and the PC. We start with writing the
Hamiltonian for the entire system. The total Hamiltonian is
\begin{eqnarray}\label{12}
   H=H_{s}+\sum_{j=0}^{N-1}(H_{pc,j}+H_{int,j}),
      \end{eqnarray}
where $H_{s}$, $H_{pc,j}$ and $H_{int,j}$ would be identified next.
\\Note that in this paper, quantum walk is defined over an
undirected graph with $N$ nodes that each node is labeled by an
integer $n\in[0,N-1]$. Also, we assume that the quantum walker has
no internal state (i.e. simple quantum walker), so that we can
describe its dynamics by a Hamiltonian of form~\cite{APS}:

\begin{eqnarray}\label{13}
\begin{array}{cc}
  H_{s}= & \displaystyle\sum_{ij}\Delta_{ij}(t)(\hat{c}^{\dag}_{i}\hat{c}_{j}+\hat{c}_{i}\hat{c}^{\dag}_{j})-
\displaystyle\sum_{j}E_{j}(t)\hat{c}^{\dag}_{j}\hat{c}_{j}, \\
  & \\
   =\hspace{-.5cm}&\displaystyle\sum_{ij}\Delta_{ij}(t)(|i\rangle\langle j|+|j\rangle\langle
   i|)+\displaystyle\sum_{j}E_{j}(t)|j\rangle\langle j|
\end{array}
     \end{eqnarray}

\begin{figure}[h] \vspace{3cm} \includegraphics{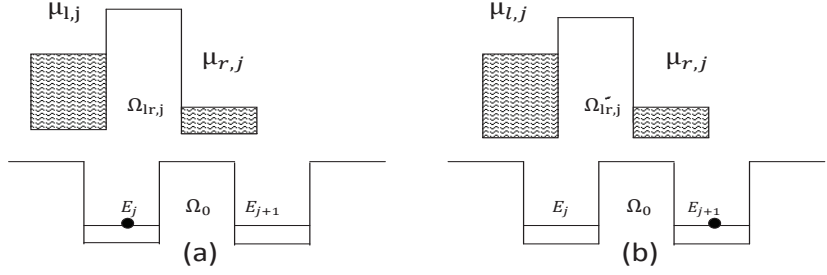} \vspace{.01cm}
\caption{Fig. 2(a) shows point contact detector $j$ monitoring the
electron in dot $j$ and Fig. 2(b) shows point contact detector $j$
when electron is placed in dot $j+1$. $\mu_{l,j}$ and $\mu_{r,j}$
are chemical potentials of left and right reservoirs of $j$th point
contact, respectively. $E_{j}$ is the on-site node energy.}
\end{figure}

where $\hat{c}^{\dagger}_{j}(\hat{c}_{j})$ is creation
(annihilation) operator and the walker at node $j$ corresponds to
the quantum state $|j\rangle=\hat{c}_{j}^{\dag}|0\rangle$. The first
term in Eq. (13) is a 'hopping' term with amplitude $\Delta_{ij}(t)$
along the edge ${ij}$ between nodes $i$ and $j$; the second term
describes 'on-site' node energies $E_{j}(t)$. We assume the hopping
amplitude between connected sites to be constant and drop on-site
energy terms (ie., ${E_{j}}=0, \forall j$). Also, for convenience,
we renormalize the time so that it becomes dimensionless~\cite{FS}.
Thus, the simple quantum walker Hamiltonian has the form:

\begin{eqnarray}\label{14}
  H_{s}=\frac{1}{4}\sum_{j=0}^{N-1}\sum_{z=1}^{l}(\hat{c}^{\dagger}_{j+z}\hat{c}_{j}+\hat{c}^{\dag}_{j}c_{j+z}).
\end{eqnarray}
The tunneling Hamiltonian $H_{pc}$ describing this system can be
written as

\begin{eqnarray}\label{15}
  H_{pc,j}=\sum_{l}E_{l,j}\hat{a}^{\dagger}_{l,j}\hat{a}_{l,j}+\sum_{r}E_{r,j}\hat{a}^{\dagger}_{r,j}\hat{a}_{r,j}+
  \sum_{l,r}\Omega_{lr,j}(\hat{a}^{\dagger}_{l,j}\hat{a}_{r,j}+\hat{a}^{\dagger}_{r,j}\hat{a}_{l,j}),
\end{eqnarray}
where $\hat{a}^{\dagger}_{l,j}(\hat{a}_{l,j})$ and
$\hat{a}^{\dagger}_{r,j}(\hat{a}_{r,j})$ are creation(annihilation)
operators in the left and right reservoirs of point contact $j$,
respectively. Also, $E_{l,j}$ and $E_{r,j}$ are the energy levels in
the left and right reservoirs of detector, and $\Omega_{lr,j}$ is
the hopping amplitude between the states $E_{l,j}$ and $E_{r,j}$. We
assume that the hopping amplitude of $j$th point contact is
$\Omega_{lr,j}$ when an electron occupies the left dot, and it is
$\acute{\Omega_{lr,j}}$ when an electron occupies the right dot.
Hence, we can represent the interaction term as

\begin{eqnarray}\label{16}
  H_{int}=\sum_{l,r}\delta\Omega_{lr,j}\hat{c}_{j}^{\dagger}\hat{c}_{j}(\hat{a}_{l,j}^{\dagger}\hat{a}_{r,j}+\hat{a}^{\dagger}_{r,j}\hat{a}_{l,j}).
\end{eqnarray}
where ($\delta\Omega_{lr,j}=\Omega_{lr,j}-\acute{\Omega_{lr,j}}$).
For simplicity, we assume that the hoping amplitudes are weakly
dependent on states $E_{l,j}$ and $E_{r,j}$, so that
$\Omega_{lr,j}=\Omega$, $\delta\Omega_{lr,j}=\delta\Omega$ and
$\mu_{l,j}(\mu_{r,j})=\mu_{l}(\mu_{r})$. Gurvitz in~\cite{AG}
applied the Bloch-type equations for a description of the entire
system with the large bias voltage($\mu_{l}-\mu_{r}$). He showed
that the appearance of decoherence leads to the collapse of the
density matrix into the statistical mixture in the course of the
measurement processes. Using Eq. (5), this analysis for our model
results in the following equation for the reduced density matrix

\begin{eqnarray}\label{17}
\begin{array}{cc}
  \frac{d}{dt}\rho_{j,k}(t)= & \frac{i}{4}[\displaystyle\sum_{z=-l}^{l}(\rho_{j,k+z}-\rho_{j+z,k})]-\Gamma(1-\delta_{j,k})\rho_{j,k}\hspace{2.4cm} \\
   &  \\
  =\hspace{-1.5cm} &
  \frac{i}{4}[\displaystyle\sum_{z=1}^{l}(\rho_{j,k+z}-\rho_{j+z,k}+\rho_{j,k-z}-\rho_{j-z,k})]-\Gamma(1-\delta_{j,k})\rho_{j,k}.
\end{array}
\end{eqnarray}

\section{Small Decoherence}
In this section, we assume that the decoherence rate $\Gamma$ is
small as ($\Gamma N\ll 1$) and consider its effect in the CTQWs on
one-dimension regular network. For this end, we make use of the
perturbation theory of linear operators ,as mentioned in~\cite{FST},
and rewrite Eq. (17) as the perturbed linear operator equation

\begin{eqnarray}\label{18}
\frac{d}{dt}\rho_{\alpha,\beta}(t)=\sum_{\mu,\nu=0}^{N-1}(iL_{(\alpha,\beta)}^{(\mu,\nu)}+U_{(\alpha,\beta)}^{(\mu,\nu)})\rho_{\mu,\nu}(t),
      \end{eqnarray}

where $\alpha,\beta,\mu,\nu$ run from 0 to $N-1$. Also,
$L_{(\alpha,\beta)}^{(\mu,\nu)}$ and
$U_{(\alpha,\beta)}^{(\mu,\nu)}$ which are the row $(\mu,\nu)$ and
column $(\alpha,\beta)$ elements of $N^{2}\times N^{2}$ matrices of
$L$ and $U$ respectively, are defined as
\begin{eqnarray}\label{19}
L_{(\alpha,\beta)}^{(\mu,\nu)}=\frac{1}{4}[\sum_{z=-l}^{l}(\delta_{\alpha,\mu}\delta_{\beta,\nu-z}-\delta_{\alpha,\mu-z}\delta_{\beta,\nu})],
      \end{eqnarray}

\begin{eqnarray}\label{20}
U_{(\alpha,\beta)}^{(\mu,\nu)}=-\Gamma\delta_{\alpha,\mu}\delta_{\beta,\nu}(1-\delta_{\alpha,\beta}).
\end{eqnarray}

Also, for our case, the initial conditions are

\begin{eqnarray}\label{21}
\rho_{\alpha,\beta}(0)=\delta_{\alpha,0}\delta_{\beta,0}.
\end{eqnarray}

Now, we want to obtain the eigenvalues and the eigenvectors of
$L+U$. For this aim, we study the perturbed eigenvalue
equation~\cite{FST}

\begin{eqnarray}\label{22}
(L+U)(V+\tilde{V})=(\lambda+\tilde{\lambda})(V+\tilde{V}),
\end{eqnarray}

that $V$ is the corresponding eigenvector with eigenvalue $\lambda$
i.e. $LV=\lambda V$. Applying first-order perturbation theory of
quantum mechanics, one can get

\begin{eqnarray}\label{23}
\tilde{\lambda}=V^{\dag}UV.
\end{eqnarray}

We assume that $\epsilon_{\lambda}$ is the eigenspace with
eigenvalue $\lambda$ and some of eigenvectors of $L$ (i.e. $\{V_{k}:
k\in I\}$) span it. For the uniform linear combination, we can
obtain $\tilde{\lambda}$ as following

\begin{eqnarray}\label{24}
\tilde{\lambda}=\sum_{k\in I}V_{j}^{\dag}UV_{k}.
\end{eqnarray}

The solution of Eq. (18) is obtained by dropping the terms
$\tilde{V}_{j}$~\cite{FST}

\begin{eqnarray}\label{25}
\rho(t)=\sum_{\lambda}e^{t(i\lambda+\tilde{\lambda})}\sum_{j\in\epsilon_{\lambda}}c_{j}V_{j}.
\end{eqnarray}

For unperturbed linear operator $L$, we have

\begin{eqnarray}\label{26}
\sum_{\mu,\nu=0}^{N-1}L_{(\alpha,\beta)}^{(\mu,\nu)}V_{(\mu,\nu)}^{(m,n)}=\lambda_{(m,n)}V_{(\alpha,\beta)}^{(m,n)},
   \end{eqnarray}

that $\lambda_{(m,n)}$ is

\begin{eqnarray}\label{27}
\lambda_{(m,n)}=\sum_{z=1}^{l}\sin(\frac{\pi
z(n+m)}{N})\sin(\frac{\pi z(m-n)}{N})
   \end{eqnarray}

and $V_{(\mu,\nu)}^{(m,n)}$ is

\begin{eqnarray}\label{28}
V_{(\mu,\nu)}^{(m,n)}=\frac{1}{N}\exp(\frac{2\pi i}{N}(m\mu+n\nu)).
   \end{eqnarray}

Using Eq. (24), one can obtain

\begin{eqnarray}\label{29}
\begin{array}{cc}
   U_{(m,n),(m',n')}& =(V^{(m,n)})^{\dag}UV^{(m',n')}\hspace{5.5cm} \\
  &  \\
  &=  -\frac{\Gamma}{N^{2}}\displaystyle\sum_{(a,b)}(1-\delta_{a,b})\exp(\frac{2\pi i}{N}[(m'-m)a+(n'-n)b]) \\
   &  \\
   &= -\Gamma\delta_{m',m}\delta_{n',n}+\frac{\Gamma}{N}\delta_{[(m'-m)+(n'-n)](mod N),0}\hspace{1.8cm}
\end{array}
   \end{eqnarray}

In what follows, we first find the degenerate eigenvalues of Eq.
(27) and then calculate the eigenvalue perturbation terms:
\\\\(a) Diagonal element($m=n$):
\\Since $U$ is diagonal over the corresponding eigenvectors, there is not such
degeneracy in our case~\cite{FST}. For these eigenvalues, the
correction terms are given by Eq. (29):
$\tilde{\lambda}_{(m,m)}=-\Gamma\frac{(N-1)}{N}$.
\\\\(b) Zero$(m+n\equiv 0 (mod N))$:
\\Since the corresponding eigenvectors can not display in the linear
combination of the initial state $\rho(0)$, this degeneracy is
irrelevant to our problem~\cite{FST}.
\\\\(c) Off-diagonal elements:
\\By Eq. (29), the off-diagonal elements are non-zero if \\$m+n\equiv m'+n'(mod
N)$.
\\To find degenerate eigenvalues satisfying the relation(c), we make divide the problem into
two separate states as follows:
\\\\\emph{The state $l=1$}: This state is equal to a cycle network, for which Eq. (27) reduces to
$\lambda_{(m,n)}=\sin(\frac{\pi (n+m)}{N})\sin(\frac{\pi
(m-n)}{N})$. $\lambda_{(m,n)}=\lambda_{(m',n')}$ results in
$\sin(\frac{\pi(m+n)}{N})=\pm \sin(\frac{\pi(m'+n')}{N})$ and
$\sin(\frac{\pi(m-n)}{N})=\pm \sin(\frac{\pi(m'-n')}{N})$. Thus, we
have
\begin{eqnarray}\label{28}\nonumber
\left\{
  \begin{array}{ll}
    m=n'+N/2, n=m'+N/2, & \hbox{for $N/2\leq m\leq N-1, N/2\leq n\leq N-1$;} \\
    m=n'-N/2, n=m'-N/2, & \hbox{for $0\leq m< N/2, 0\leq n< N/2$;} \\
    m=n'+N/2, n=m'-N/2, & \hbox{for $N/2\leq m\leq N-1, 0\leq n< N/2$;} \\
    m=n'-N/2, n=m'+N/2, & \hbox{for $0\leq m< N/2, N/2\leq n\leq N-1$.}
  \end{array}
\right.
 \end{eqnarray}
The correction terms to these eigenvalues are
$\tilde{\lambda}_{(m,n)}=-\Gamma\frac{(N-2)}{N}$.
\\\\\emph{The state $l\geq 2$}: In this case, the structure of cycle network can be destroyed
 by $2l$ additional bonds in the network.
\\$\lambda_{(m,n)}=\lambda_{(m',n')}$ implies to
\\$\displaystyle\sum_{z=1}^{l}\sin(\frac{\pi
z(n+m)}{N})\sin(\frac{\pi
z(m-n)}{N})=\displaystyle\sum_{z=1}^{l}\sin(\frac{\pi
z(n'+m')}{N})\sin(\frac{\pi z(m'-n')}{N})$.
\\\\One can see that there is not any degeneracy for this
state. Thus, correction terms to these eigenvalues are
$\tilde{\lambda}_{m,n}=-\Gamma\frac{(N-1)}{N}$.
\\\\The mixing time bound for the state $l=1$ was provided in ~\cite{FST, VKR}. \\In the
following, we focus on the state $l\geq 2$(one-dimension regular
network under condition $l\geq 2$) and in the end, compare our
results with~\cite{FST, VKR}'s results for cycle.
\\Based on the above analysis, Eq. (25) can be written as

\begin{eqnarray}\label{31}
\rho(t)=\sum_{(m,n)}\frac{1}{N}(\delta_{m+n,0}+\delta_{m+n,N})V^{(m,n)}
   \end{eqnarray}

and from Eq. (21), we have

\begin{eqnarray}\label{32}
\rho(0)=\sum_{(m,n)}\frac{1}{N}V^{(m,n)}.
   \end{eqnarray}

Thus, the full solution is
\begin{eqnarray}\label{33}
\begin{array}{cc}
  \rho_{\alpha,\beta}(t)&=\frac{\delta_{\alpha,\beta}}{N}+\frac{1}{N^{2}}\displaystyle\sum_{(m,n)}(1-\delta_{[m+n](mod
N),0})e^{t(i\lambda_{(m,n)}+\tilde{\lambda}_{(m,n)})} \\
   &  \\
   & \times\exp[\frac{2\pi i}{N}(m\alpha+n\beta)].\hspace{5.5cm}
\end{array}
   \end{eqnarray}

The probability distribution $P_{j}(t)$ of the quantum walk is
specified by the diagonal elements of the reduced density matrix,
i.e.

\begin{eqnarray}\label{34}
\begin{array}{cc}
  P_{j}(t) & =\frac{1}{N}+\frac{1}{N^{2}}\displaystyle\sum_{(m,n)}(1-\delta_{[m+n](mod
N),0})\times[e^{-\Gamma\frac{N-1}{N}t}]\hspace{2.9cm}\\
   &  \\
  & \times\exp{[it\displaystyle\sum_{z=1}^{l}\sin(\frac{\pi
z(m+n)}{N})\sin(\frac{\pi z(m-n)}{N})]}\times\exp[\frac{2\pi
i}{N}(m+n)j].
\end{array}
   \end{eqnarray}

\section{Mixing time}
There are two distinct notions of mixing time for quantum walks in
the literature:
\\\emph{Instantaneous mixing time}: Instantaneous mixing time is
defined as the first time instant at which the probability
distribution of the walker's position is $\epsilon$-close to the
uniform distribution~\cite{DHSS}. Thus, the instantaneous mixing
time is
\begin{eqnarray}\label{35}
M_{inst,\epsilon}=min\{t:\|P_{j}(t)-\frac{1}{N}\|_{tv}<\epsilon\},
   \end{eqnarray}
where here we use the total variation distance to measure the
distance between two distributions $P,Q$:
$\|P-Q\|_{tv}=\sum_{i}|P(i)-Q(i)|$.
\\Now by Eq. (34), we calculate an upper bound on the
$M_{inst,\epsilon}$
\begin{eqnarray}\label{36}
\begin{array}{cc}
  |P_{j}(t)-\frac{1}{N}| &
  =e^{-\Gamma\frac{N-1}{N}t}|\frac{1}{N^{2}}\displaystyle\sum_{(m,n)}
(\exp[{it\displaystyle\sum_{z=1}^{l}\sin(\frac{\pi
z(m+n)}{N})\sin(\frac{\pi
z(m-n)}{N})}] \\
   &  \\
   &\times\exp[\frac{2\pi i}{N}(m+n)j])-\frac{1}{N}|.\hspace{5.5cm}
\end{array}
   \end{eqnarray}
\\
\\
By noting that \\$\exp[it\displaystyle\sum_{z=1}^{l}\sin(\frac{\pi
z(m+n)}{N})\sin(\frac{\pi z(m-n)}{N})]\times\exp[\frac{2\pi
i}{N}(m+n)j]\leq1$,
\\\\ one can obtain
\begin{eqnarray}\label{37}
|P_{j}(t)-\frac{1}{N}|\leq e^{-\Gamma\frac{N-1}{N}t}.
   \end{eqnarray}
Thus, we have

\begin{eqnarray}\label{38}
\sum_{j=0}^{N-1}|P_{j}(t)-\frac{1}{N}|\leq
Ne^{-\Gamma\frac{N-1}{N}t}.
   \end{eqnarray}
Based on the above definition, the upper bound of instantaneous
mixing time can be obtained in the following way:

\begin{eqnarray}\label{39}
Ne^{-\Gamma\frac{N-1}{N}t}\leq\epsilon,
   \end{eqnarray}

and therefore,

\begin{eqnarray}\label{40}
M_{inst,\epsilon}\leq\frac{1}{\Gamma}\ln(\frac{N}{\epsilon})[1+\frac{1}{N-1}].
   \end{eqnarray}

The $M_{inst,\epsilon}$ for cycles was provided in~\cite{FST}, that
is

\begin{eqnarray}\label{40}
M_{inst,\epsilon}\leq\frac{1}{\Gamma}\ln(\frac{N}{\epsilon})[1+\frac{2}{N-2}].
   \end{eqnarray}

These relations show that the instantaneous mixing time for
one-dimension regular network with distance parameter $l(l\geq 2)$
is shorter than the one for cycle network. Also, this quantity is
independent of distance parameter $l$ .
\\
\\\emph{Average mixing time:} To define the notion of
average mixing time of CTQWs, we use the time-averaged probability
distribution, i.e.
$\bar{P}_{j}(t)=\frac{1}{\tau}\int_{0}^{\tau}P_{j}(t)dt$. The
average mixing time measures the number of time steps required for
the time-averaged probability distribution to be $\epsilon$-close to
the limiting distribution~\cite{AAKV}, i.e.
\begin{eqnarray}\label{40}
M_{ave,\epsilon}=min\{t|\forall
\tau>T:\|\bar{P}_{j}(t)-\frac{1}{N}\|_{tv}<\epsilon\}.
   \end{eqnarray}
In the following, we want to obtain the lower bound of average
mixing time for one-dimension regular network$(l\geq2)$.
\\Applying Eq. (36) for large $N\gg1$, we have
\begin{eqnarray}\label{40}
\|\bar{P}_{j}(t)-\frac{1}{N}\|\leq|\frac{1}{T}\int_{0}^{T}(e^{-\Gamma
t}+\frac{1}{N})dt-\frac{1}{N}|=\frac{1}{\Gamma T}[1-e^{-\Gamma T}].
   \end{eqnarray}
Summing over $j$ to calculate the total variation distance, we have
\begin{eqnarray}\label{40}
\frac{N}{\Gamma T}(1-e^{-\Gamma T})\leq \epsilon.
   \end{eqnarray}
Then we assume that $\Gamma T\gg 1$ (since $N\gg 1$ and $\Gamma N\ll
1$, thus $T\gg N$) and achieve
\begin{eqnarray}\label{40}
M_{ave,\epsilon}\geq\frac{N}{\Gamma\epsilon},
   \end{eqnarray}
which is similar to the average mixing time bound produced for cycle
in~\cite{VKR}.

 {\bf{Physical interpretation:}}\\
As mentioned in Sec. 3, we assumed the hopping amplitude between all
of connected sites to be equal. Also, we supposed that every
point-contact detector which is coupled to states localized on the
corresponding node, measures the position of the particle in space
of graph. Since these detectors identify the path of the walker
took, quantum interference which is the result of an uncertainty in
the path, is then lost and therefore the mixing time is independent
of distance parameter $l$~\cite{NVPC}. However, one notes that this
explanation might not be broadly accepted.
\section{Conclusion}
We studied the effect of small decoherence on one-dimension ring
lattice of $N$ nodes in which every node is linked to its $2l$
nearest neighbors ($l\geq 2$ on either side). In our investigation,
this network was represented by the system of identical
tunnel-coupled quantum dots. As the detector, we used the point
contact in close proximity to one of the dots. For description of
the entire system, we applied the Bloch-type equations. Then, we
calculated the probability distribution and the mixing time bound.
We showed that the mixing time bound is independent of parameter
$l(l\geq 2)$. Also, we observed that this quantity is inversely
proportional to rate of decoherence, as mentioned in~\cite{FST,
VKR}. Hence, decoherence can make short the mixing time on these
networks. Moreover, we found that the instantaneous mixing time for
one-dimension regular network is smaller than the one for cycle
network.

\end{document}